# On the Mechanical and Electronic Properties of Thiolated Gold Nanocrystals

K. Smaali,[a,λ] S. Desbief,[a,λ] G. Foti,[b,c] T. Frederiksen,[c,d] D. Sanchez-Portal,[b,c] A. Arnau,[b,c,e] J.P. Nys,[a] P. Leclère,[f] D. Vuillaume[a] and N. Clément[a*]

We present a quantitative exploration, combining experiment and simulation, of the mechanical and electronic properties, as well as the modifications induced by an alkylthiolated coating, at the single NP level. We determine the response of the NPs to external pressure in a controlled manner by using an atomic force microscope tip. We find a strong reduction of their Young modulus, as compared to bulk gold, and a significant influence of strain in the electronic properties of the alkylthiolated NPs. Electron transport measurements of tiny molecular junctions (NP/alkylthiol/CAFM tip) show that the effective tunnelling barrier through the adsorbed monolayer strongly decreases with increasing the applied load, which translates in a remarkable and unprecedented increase of the tunnel current. These observations are successfully explained using simulations based on finite element analysis (FEA) and first-principles calculations that permit to consider the coupling between the mechanical response of the system and the electric dipole variations at the interface.

## A Introduction

Central to the success of virtually all applications of NPs[1,2] is the need to tailor their properties with organic coatings, often self-assembled monolayers (SAMs), which impact both stability and specific functionality.[3] Substantial effort has focused on optimizing the activity of the immobilized layer of, for example, antibodies or enzymes. Nevertheless, only recently has the influence of the SAM on the properties of the underlying NP been studied. It has been found that the covering layer might have a remarkable impact in the NP structure. In the case of thiolated molecules, instead of a sharp gold-molecule boundary, a 0.25-nm-thick interfacial shell was found to contain enlarged Au-Au distances and an interpenetration of the thiol ligand species.[4,5] Another study revealed that the use of a simple propane thiol monolayer on a nanocrystal was enough to modify its facets.[6]

A large number of questions still remain regarding both the electric and mechanical properties of gold NPs, with and without organic coatings, at the single-NP level. For example, it is still unclear whether the mechanical properties of the NPs are comparable to those of bulk[6,7] and how they are modified by the presence of organic coatings. Furthermore, although the tuning of the electronic properties of Au NP by functionalization has been demonstrated,[8,9] the impact of the strain of the covering layer in those changes remains largely unexplored at a quantitative level. Here we present an example of such a quantitative analysis. Our measurements are based on a recently developed technique to grow NP with an organic coating on only one side and an ohmic contact on the other,[10] a powerful test-bed for molecular electronics.[11] This allows us to address the mechanical and electronic properties of thiolated nanocrystals, by using an array of 10-nm facetted nanocrystals with an ohmic bottom contact and a top contact made by an atomic force microscope (AFM) tip at an adjustable loading force. Contrary to the case of extended SAMs on flat surfaces, the use of nano-SAMs (lateral dimensions on the nanoscale) makes the contact area independent of the applied force. Therefore, with this set-up it is possible to measure the load dependence of the mechanic and electric properties with high precision.

With the fabrication technique reported in Reference 10 (see Methods), half of the nanocrystal is buried in highly doped silicon (Fig. 1a). Its structure is very close to that of an ideal cuboctahedron or truncated octahedron NP (Fig.1b).[12] A scanning tunneling microscope (STM) image shows a flat top surface (Fig.1c) and scanning transmission electron microscopy (STEM) clearly reveals the facets (Fig. 1d) for this NP.[10] This structure is ideal for testing the mechanical properties of naked gold nanocrystals or molecularly functionalized nanocrystals with an AFM tip, for several reasons. First, the force applied with the tip can be precisely tuned. Second, the tip curvature radius (~40 nm) is much larger than the flat top surface of the nanocrystal (<10 nm); thus, this system can be considered as an ideal parallel-plate scenario. Third, statistical analysis can easily be performed by using an array with a large number (typically several thousands) of nanocrystals. Finally, due to the ohmic contact between the nanocrystal and highly doped silicon, the electronic and mechanical properties of the SAM coating can be measured simultaneously with a conducting AFM (CAFM) tip (Fig.1e).

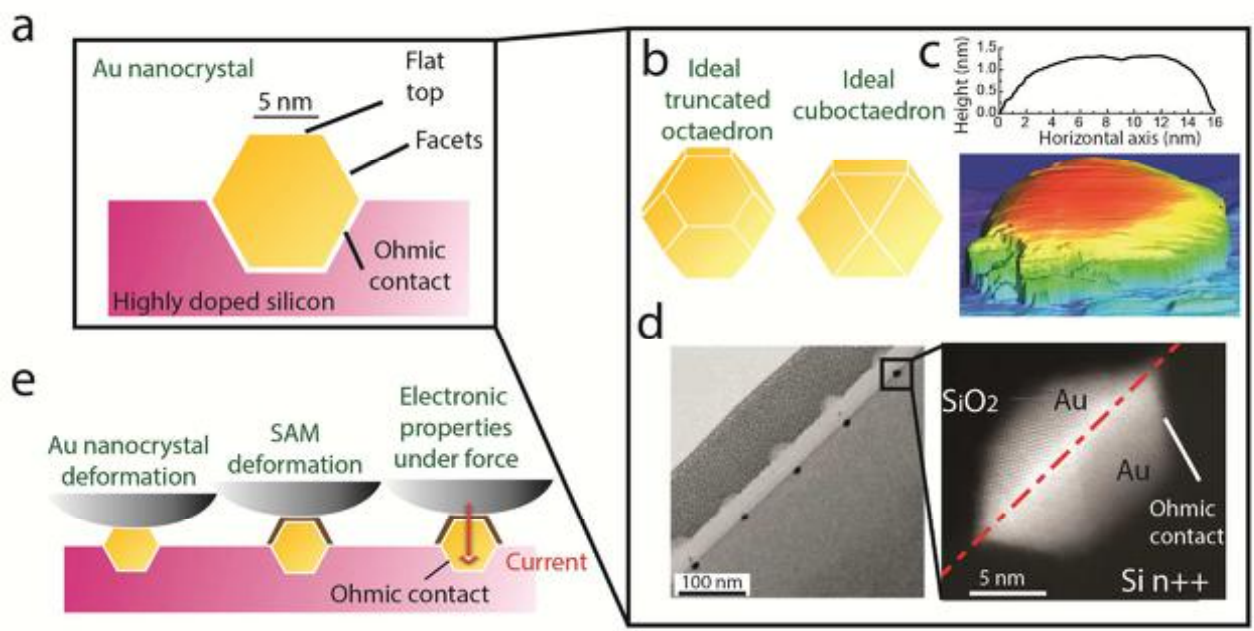

**Figure 1 Gold nanocrystals description and experimental setup.**
a, Schematic cut-view of the facetted gold nanocrystal. Thanks to the unique fabrication technique (see methods), these nanocrystals are well attached to a highly doped substrate with an ohmic contact.
b, Fabricated gold nanocrystals resemble to either ideal truncated octahedron or cuboctahedron NPs. Annealing temperature (260°C) was very close to the predicted temperature to obtain these ideal NPs.[12]
c, Scanning Tunneling Microscope (STM) image obtained on a single Au nanocrystal. The cut line shows a flat top (roughness < 2Å). The green step observed in the STM image is linked to the HF etching prior to imaging in Ultra-High Vacuum (UHV) (partial consumption of highly doped silicon).
d, STEM images taken from ref. 10 on five (left) and single (right) Au nanocrystals. Atoms and facets of the nanocrystal are clearly seen.
e, Schematic view of the experimental setups. From left to right: peak-force AFM on an uncoated nanocrystal, on a alkylthiol-coated nanocrystal and CAFM on a coated nanocrystal.

In this work, we show that the Young modulus of tiny single crystal NPs (<8 nm in diameter) buried in silicon substrate is ~20 GPa, smaller than the one of free-standing NPs (~40 nm). We also estimate that the Young modulus of alkylthiol monolayers, self-assembled on top of them, is in the range 0.5-2.8 GPa. Electron transport measurements of tiny molecular junctions made with NPs by self-assembled alkylthiol monolayers (chain length from 8 to 18 carbon atoms) reveal unprecedented behaviors: i) a strong decrease of the tunnel current decay factor $\beta$ from 0.9 to 0.2 per carbon atoms when the loading force is increased only up to 30 nN, ii) a decrease by ~0.4 eV of the HOMO level with respect of the Au Fermi energy. These results are well explained by a force-induced modification of the Au-alkylthiol interface dipole, and supported by DFT calculations.

## B Elastic properties

A Young's Modulus of a gold NP

*Peak-force AFM experimental study*

The mechanical properties are obtained by direct measurement of the deformation with a peak-force Atomic Force Microscope (AFM, Brüker©, see Methods) on an array of nanocrystals at a given load (peak force) of 150 nN (Fig. 2a). This force generates sufficient deformation of the nanocrystal for quantitative study, but is still in the elastic deformation regime. The inset in Fig. 2b shows a zoom on three nanocrystals. Each nanocrystal exhibits a bright ring that indicates a large deformation when the AFM tip is on the facets. When the maximum deformation is measured on the facets and at the center of the top surface for each nanocrystal in a large array of 686 nanocrystals, the constructed deformation histogram reveals two peaks (Fig. 2b) corresponding to top and facets. The deformation value on the facets (~3 nm on average) seems large, given the dimensions of the nanocrystal (height of about 2–3 nm; see Fig. 1d). However, one should take into account that peak force deformation measurements are only relevant along the vertical axis and, therefore, the deformations of tilted facets cannot be obtained reliably (see Supplementary Information [SI], Fig. S1). When the tip is on top of the nanocrystal, the total deformation (tip and nanodot) of 0.93 ± 0.08 nm is found experimentally (Fig.2b). Below, we use FEA to estimate NPs Young modulus.

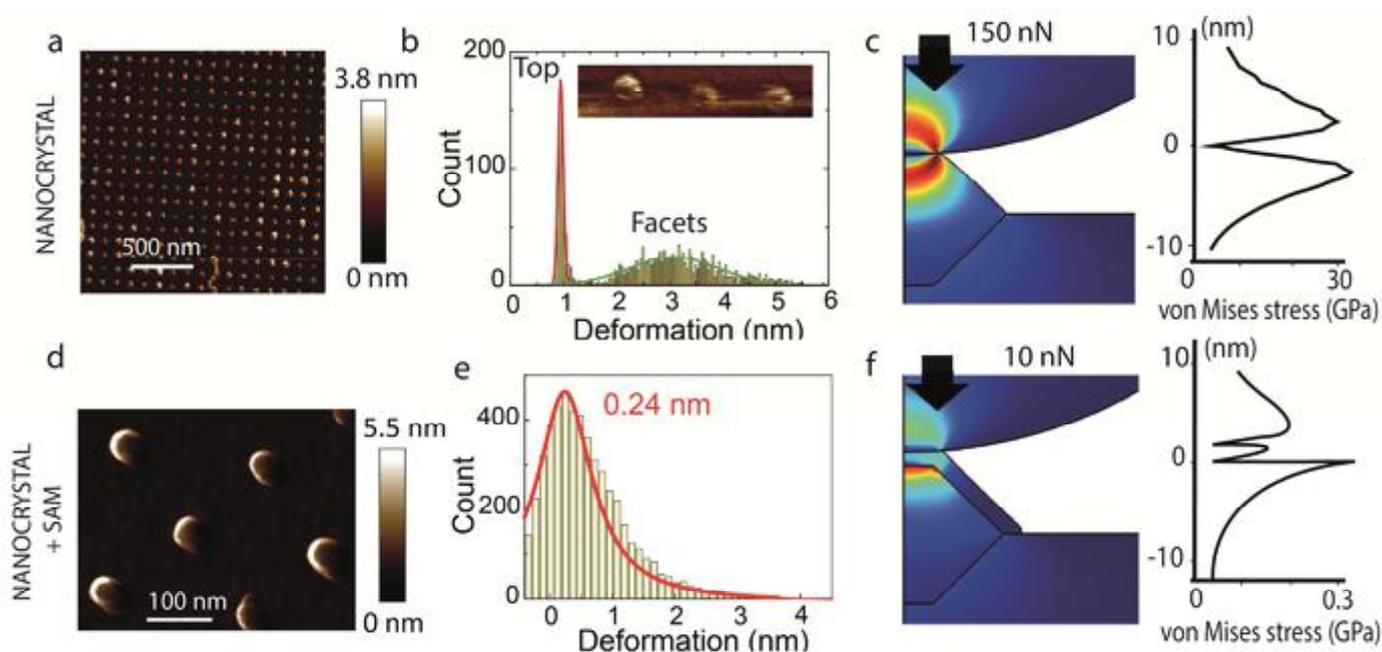

**Figure 2 Elastic properties of uncoated and coated Au nanocrystals.**
a, Peak-force AFM image of the measured deformation on an array of uncoated Au nanocrystals.
b, Histograms of deformation related to top and facets of Au nanocrystals. Inset: zoom on three Au nanocrystals.
c, 2D-FEA von Mises stress map of the tip indentation at 150 nN in the Au nanocrystal. NP Young's modulus $E_{NP}$~20 GPa has been selected to get a total deformation of 0.93 nm. This structure, considered as "ideal" might differ from the experimental structure such as the presence of air gap at the gold/Si interface, however the STEM image in Fig.1c present a close-to-ideal structure at atomic resolution. Right: cut view of the von Mises stress along the symmetry axis.
d, Peak-force AFM deformation image of a $C_{12}$-alkylthiolated Au nanocrystal
e, Histograms of deformation related to the top of coated nanocrystals.
f, 2D-FEA von-Mises stress map of the tip indentation in the SAM at 10 nN. SAM's Young modulus $E_{SAM}$~1.4 GPa has been tuned to get a total deformation of 0.24 nm ($E_{NP}$=20 GPa).

*Finite Element Analysis of gold nanocrystals Elastic properties*

The experimentally measured deformation is correctly simulated (Fig. 2c) by FEA if we assume a Young's modulus $E_{NP}$ of 20 ± 2 GPa. We considered the full structure including the silicon substrate in which the gold NP is half buried and found a negligible deformation of the substrate (see Fig.S1). For FEA, we have selected to show the von Mises stress, often used in determining whether an isotropic and ductile metal will yield when subjected to a complex loading condition. It has the advantage to clearly delimit each material on the images which is useful for SAM deformation estimation and to highlight stressed regions. The von Mises stress is equally distributed at both sides of the contact, with a maximum located 4 nm from the contact. The obtained Young modulus value is lower than the bulk value that is usually considered for $E_{NP}$ (74-80 GPa).[6,7] Below, we discuss the low Elastic modulus estimated for sub-10 nm gold nanocrystals.

*Discussion on the 20 GPa gold nanocrystal Elastic modulus*

Theoretical studies suggest that Young's modulus for spheric NPs can be reduced by up to 50% from the bulk value.[13] Previous reports on the elastic properties of gold nanocrystals obtained by time-resolved spectroscopy (mainly nanorods of 20 nm diameter and 100 nm length) have reported either a bulk value[14] (~79 GPa) or lower[15] (~64 GPa). Very recently, it was shown that the Young's modulus is reduced (to ~40 GPa) when the cylindrical symmetry of these nanorods is ruptured by the presence of facets and that this effect is amplified as the length of the nanorod is reduced.[16] This effect was explained by the anisotropic elastic properties of single-crystal nanorods and heating effect of optical-induced plasmonic resonance. Our facetted gold NPs are also single-crystal NPs. As a consequence, the elastic modulus is likely anisotropic with values down to 42 GPa. But even more importantly, we observed microtwins with {111} twin boundaries.[10] Twins likely reduce the elastic modulus of our single-crystal NPs.[17,18] It remains that the large surface to volume ratio may further lower the measured NP elastic modulus.

B Young's modulus of SAMs covering gold NPs

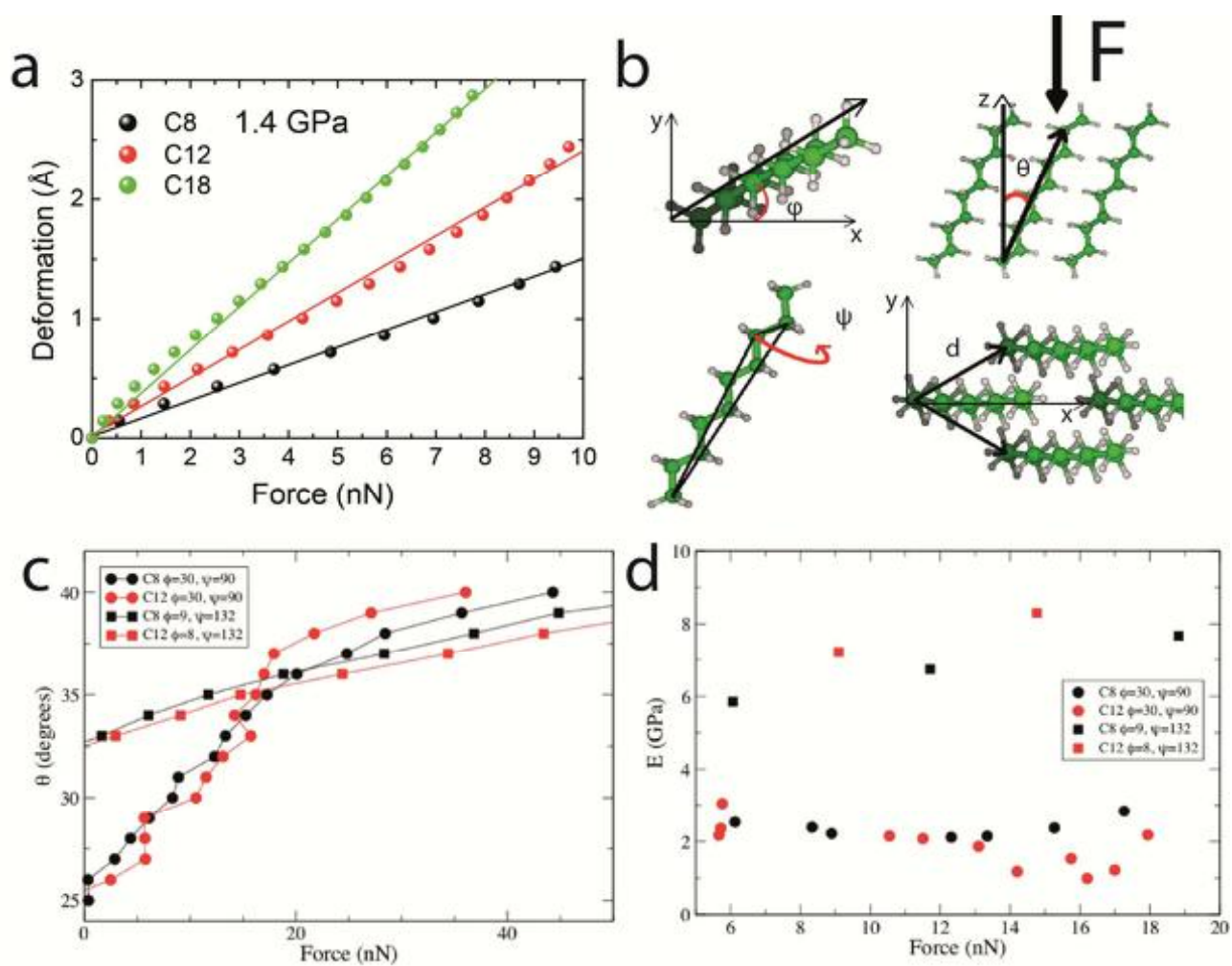

**Figure 3 Theoretical study of Alkyl SAM elastic properties**
a, FEA corresponding to Fig. 2f for 3 different alkyl chain lengths ($C_8$, $C_{12}$, $C_{18}$) at $E_{SAM}$=1.4 GPa. A linear deformation is not observed with a planar SAM (i.e., a SAM with a large lateral extension grafted on an Au surface), in which, due to the tip indentation into the SAM, the contact area increases with force (supplementary, Fig.S2).
b, Schematic view of the SAM representation. The molecular tilt angle θ is the angle between the molecular principal axis and the surface normal while Φ and Ψ represent the rotational and twist angle, respectively. The parameter d is the intermolecular distance.
c, Tilt versus force for two possible configurations of the $C_8$ and $C_{12}$ SAM estimated from DFT calculations.
d, Estimated Young's modulus E for $C_8$ and $C_{12}$ as a function of the applied force from DFT calculations.

*Peak-force AFM experimental study*

Fig. 2d shows the measured deformation, and Fig. 2e the related histograms when a dodecanethiol monolayer (archetype of a tunnel barrier in molecular junctions[19-24] with 12 carbon atoms) is chemically grafted on the nanocrystals (see Methods). A mean deformation of 0.24 nm is obtained under a pressure of 10 nN (the load is reduced to keep the SAM deformation in the elastic limit). The dispersion (1 nm at half peak) is larger but in the range of previously reported deformation dispersion for laterally extended SAMs (0.3 nm at half peak for same monolayer and same deformation).[26] Difficulty of precise deformation measurements in the Å range on such thin layers coating nanocrystals may be at the origin of larger measured dispersion. Below we consider the full width half maximum deformation to estimate the Young modulus of SAMs coating gold NPs by FEA.

*Finite Element Analysis and discussion on SAM coated gold nanocrystals*

Based on FEA with $E_{SAM}$ the SAM Young modulus, we get 0.5<$E_{SAM}$<2.8 GPa. $E_{NP}$=20 GPa is considered as reported above to take into account the deformation of the NP below the SAM (considering $E_{NP}$=78 GPa, as previously reported, would only affect $E_{SAM}$ by 2%). Peak-force AFM images indicate that adhesion is almost cancelled on alkylthiolated nanocrystals, in agreement with expected contrast of hydrophilicity/hydrophobicity of $SiO_2$/alkyl SAM (supplementary, Fig. S3). This structure prevents parasitic role of hydration on the electronic properties of alkylthiolated gold nanocrystals.[25] A similar result is obtained for octanethiol molecules. The FEA simulation reveals a large stress at the monolayer boundary with the NPs (Fig. 2f), the importance of which will be further discussed below. The Young's modulus extracted here for a SAM with 12 methyl groups on a gold nanocrystal is lower than the $E_{SAM}$ of ~4 GPa estimated for close-packed alkylthiol-functionalized NP arrays[6] (with the assumption of an $E_{NP}$ similar to that of bulk), but is on a par with results by Del Rio et al.[26] for SAMs on a planar substrate and by Callister et al. for polyethylene.[27] Supposing the same "average" Young's modulus $E_{SAM}$ of 1.4 GPa for different alkyl chain lengths $C_N$, where $N$ is the number of carbon atoms, the SAM deformation $\Delta d$ as a function of the loading force obtained by FEA is plotted in Fig. 3a. The observed dependence follows closely that of Hooke's parallel-plate formula:

$$\Delta d(N) \approx \frac{d_0(N)F}{SE_{SAM}} \quad (1)$$

where $F$ is the applied force, $S$ is the contact area (~55 nm²), and $d_0(8)=1$nm, $d_0(12)=1.5$ nm, $d_0(18)= 2.5$ nm are the theoretical lengths of alkyl chains in their all-trans configuration at zero force, that also agree with the average experimental thicknesses measured by AFM (see SI, Fig.S4).

Below, we use first-principles density functional theory (DFT) simulations to complement our FEA approach and obtain a molecular-level understanding of the behavior of the monolayer ($C_8$ and $C_{12}$) under an external force.

*First principles density functional theory*

We study the effect of the applied force, depending on the length and orientation of the molecules in the SAM, excluding the effects of the substrate and anchoring groups. The orientation of molecules in the monolayer is defined by the three angles (Fig.3b): tilt ($\theta$), rotational ($\Phi$) and twist ($\psi$). A $\sqrt{3} \times \sqrt{3}$ 30° lattice geometry is assumed for the molecular organization in the SAM on the <111> oriented Au top surface of the nanodots,[9] with an experimental intermolecular distance $d$ of 5.05Å. Electronic structure calculations and relaxations are described in the Methods and SI (Fig. S5 and S6).

Fig. 3c shows the variation of the tilt angle under strain for several configurations corresponding to different values of $\Phi$ and $\psi$. As explained in detail in the SI we tilt the molecules as rigid rods using different configurations. This allows obtaining a smooth behavior of the energy versus tilt angle that can be numerically differentiated. Although the deformation properties barely depend on the chain length (i.e., the tilt angle-force dependence is the same for the $C_8$ and $C_{12}$ molecules), they strongly depend on ($\Phi$, $\psi$) as shown in the SI. This highlights the dependence of the calculated elastic constants on the detailed structure of the layer, which agrees with the large variance in the measured Young's modulus. An initial analysis of the energy landscape for a tilt angle $\theta=30°$, close to the equilibrium value at zero load, reveals the minimum of energy around (8°, 132°) for both molecules. Using this configuration we obtain an estimation of the Young's modulus for applied load in the range 0-20 nN of $E_{SAM}$~7 GPa [our definition of the Young's modulus for finite deformation can be found in Eq. (S1) of the SI]. This value is considerably larger than our experimental estimation. However, we must take into account that our estimations provide upper limits for the Young modulus of the layer. Indeed, it is easy to find starting configurations that are not far in energy and give rise to "softer" layers. Under applied stress these "softer" configurations will be the most relevant, since configurations with large Young's modulus will be rapidly destabilized as a function of the applied stress (since its energy increases faster). For example, for the (30°, 90°) configuration we estimate a $E_{SAM}$~2 GPa for both $C_8$ and $C_{12}$ molecules, which is fairly independent of the applied force as shown in Fig.3d. This result is close to the experimental and FEA values. The fact that the values of $E_{SAM}$ estimated by first-principles DFT calculations using plausible monolayer configurations are close to those obtained from the experiments can be interpreted as a validation of the FEA approach. In the electronic properties section below, we consider the SAMs as homogeneous films with a constant Young's modulus.

## C Electronic properties

*Current histograms generated by CAFM on a large array of thiolated gold NPs*

At a given load and bias, the electronic properties of the SAM can be investigated by Conducting AFM (CAFM, Fig. 4a, inset and Methods). The bridging of metal electrodes by alkanes (simple saturated carbon chains) is used as a prototype tunnel junction for investigating electronic and transport properties across molecule-electrode interfaces.[20-23,29,30] Alkanes have a large energy gap (of several eV) between the highest occupied and the lowest unoccupied molecular orbital (HOMO-LUMO gap).[21,31-33] Alkane junctions display typical off-resonance transport characteristics as the Fermi energy $E_F$ of the metal electrodes falls into the insulating HOMO-LUMO gap. The low-bias tunnelling probability of electrons can be understood in terms of an energy barrier $\varphi$, related to the position and alignment of the molecular level with respect to $E_F$, and a tunneling length set by the number of carbon atoms in the molecular backbone. Because alkylthiolated nanocrystals have a lower resistivity than native silicon oxide, they are clearly distinguished in the CAFM image (Fig.4a).[11] Due to the linear scale, bright spots mainly correspond to high conductance junctions. After thiol adsorption and cleaning in an ultrasonic bath, typically 80-85% of the dots are still there (see SI, Fig.S3 and S11). The 15-20% remaining dots are sometimes lying on the surface but do not respond electrically because of the presence of a native oxide layer. As a consequence, they are not considered in the statistical study.

Histograms of the current are generated from the CAFM image, with one count per nanocrystal. Histograms for $C_8$ at 3, 7.5 and 30 nN are shown in Fig.4b. Two peaks of conductance, fitted by two log-normal distributions (see SI, Fig. S7), can be observed. They are attributed to different molecular organization phases in the SAM,[11,14] and tend to merge when the applied force is increased. The distribution is rather large but in the same range as the one typically observed in single-molecule break junction experiments.[35] The mean current level increases up to several orders of magnitude with a load of 3 to 30 nN (Fig. 4b and 4c).

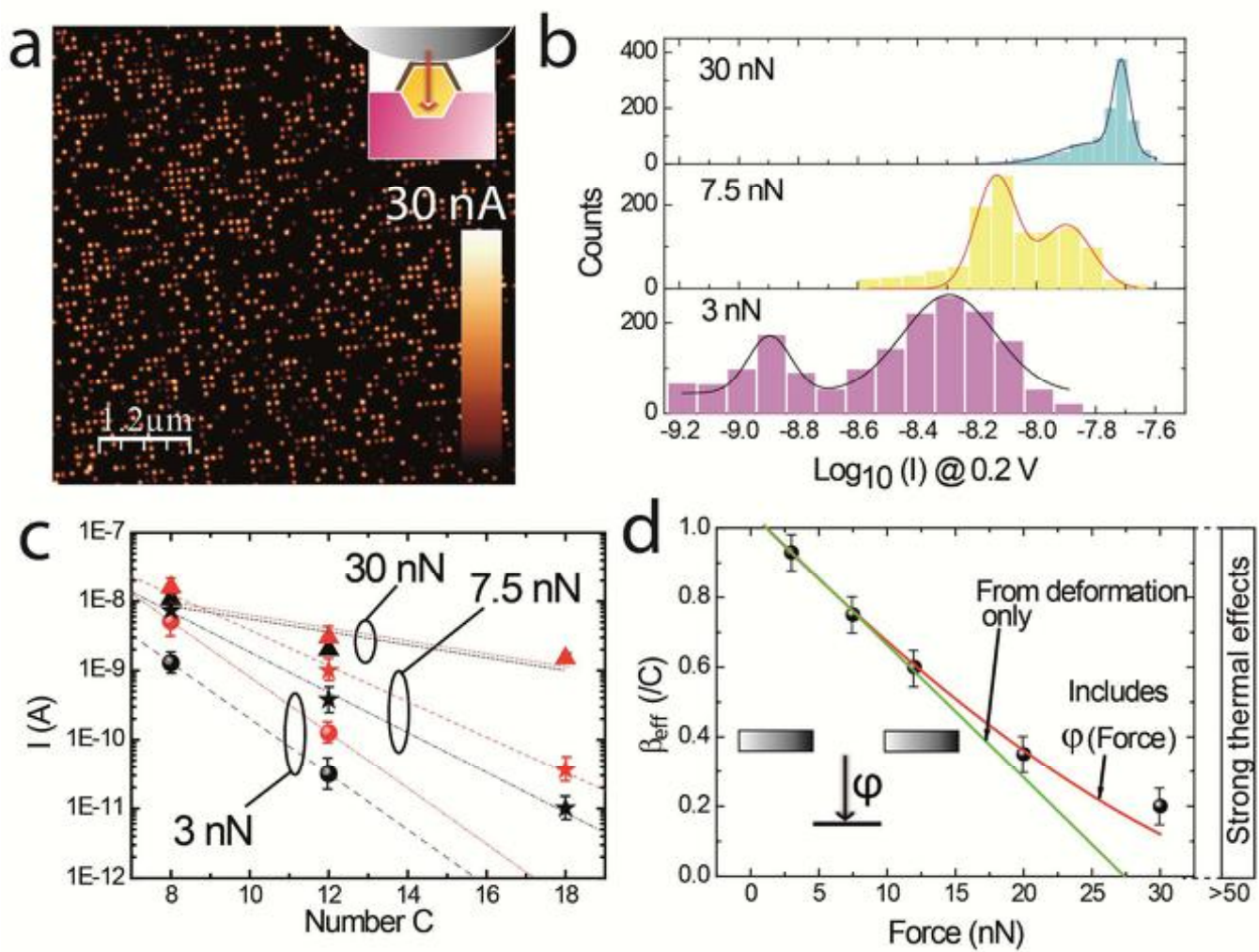

**Figure 4 Electronic properties of Alkylthiolated gold nanocrystals**
a, CAFM image of $C_8$-coated Au nanocrystals at 0.2 V and 30 nN. Inset: schematic view of the experimental setup.
b, Current histograms for $C_8$-coated nanocrystals at 0.2 V and 3 different loads. The number of counts is about 1600 per histogram.
c, Current vs number of carbon atoms for $C_8$, $C_{12}$, $C_{18}$ at 3 different loads. Each point corresponds to the maximum of a peak in current histograms. Two colors (black and red) are used to distinguish both peaks.
d, $\beta_{eff}$ obtained from c, is plotted as a function of force. Inset: band diagram showing the potential φ between the HOMO and Fermi level of the electrodes. The βeff decreases rapidly with increasing force from 0 to 30 nN (Fig. 4d) and above 50 nN, Joule-induced heating are observed on C8 SAMs, leading to nanocrystal sublimation (supplementary, Fig. S6).

*Discussion on tunnel decay rate*

If we assume a model where the SAM acts as a tunneling barrier, then the conductance can be defined as $G=Ae^{-\beta_{eff}N}$ where $A$ is the contact conductance and $N$ the number of carbon atoms. The decay constant $\beta_{eff}$, extracted by fitting the current $I$ vs. $N$ log-lin plots shown in Fig.4c supposing $A$ constant,[35-38] is a parameter including both the variation of the tunnel barrier height and the tunnel distance (i.e., SAM thickness) with the applied force. $\beta_{eff}$ is not strictly equal for the two conductance peaks, but for simplicity, we considered a single $\beta_{eff}$ for both peaks with an error bar (Fig.4d). $\beta_{eff}$ decreases rapidly with increasing force from 0 to 30 nN and above 50 nN, Joule-induced heating is observed on $C_8$ SAMs, leading to nanocrystal sublimation (see SI, Fig. S8). As $\beta_{eff}$ tends to zero, the measured current level barely depends on the number of carbon atoms in the monolayer. For comparison, in planar alkyl SAMs, $\beta_{eff}$ remains above 0.75.[41] The most representative effect is the current level for $C_{18}$, which is negligible at 3 nN and becomes similar to that of $C_8$ at 30 nN. This finding is partly

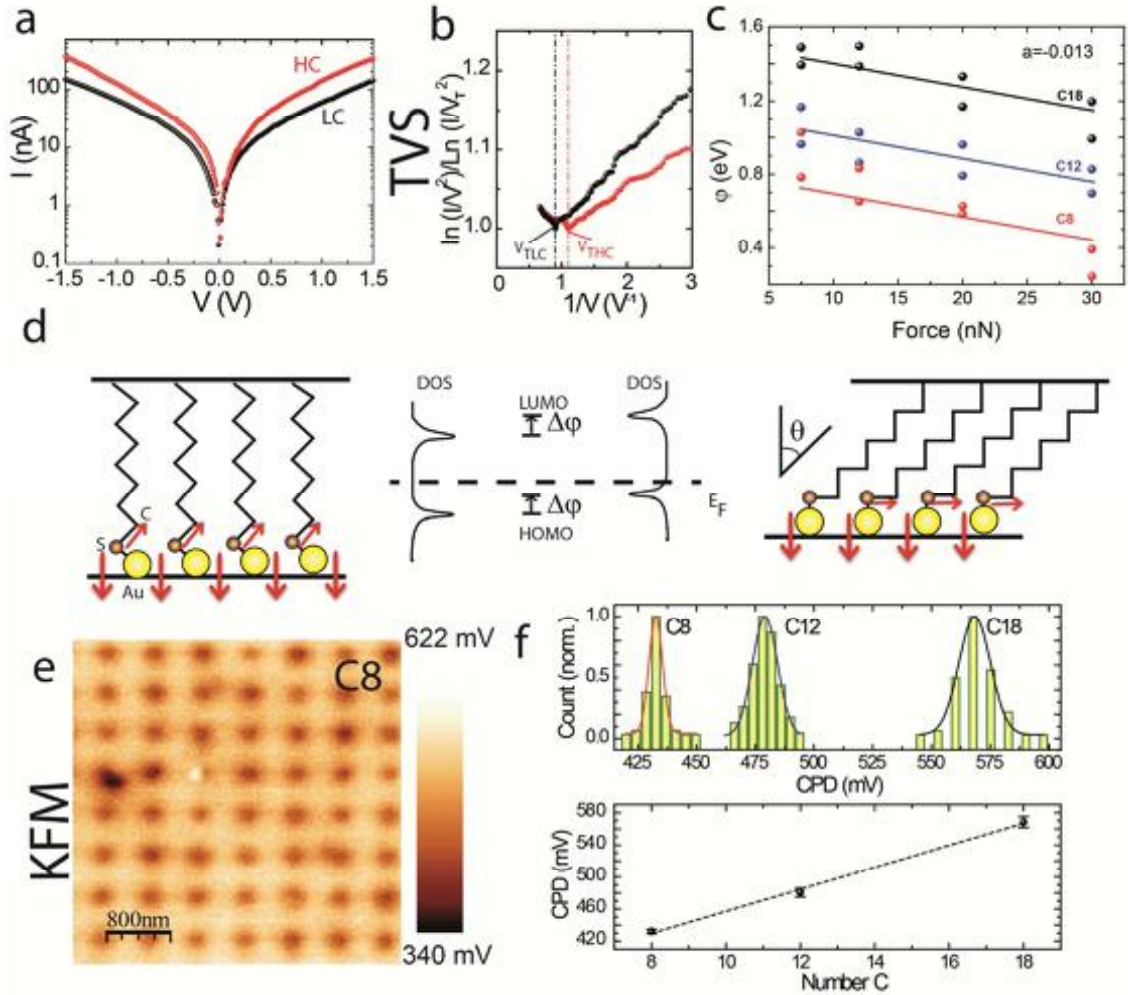

**Figure 5: Molecular orbital position of Alkylthiolated gold nanocrystals**
a, *I-V* for the two representative curves of the high and low conductance peaks for $C_8$ molecules.
b, Normalized TVS obtained from (a) used to extract minima $V_{TLC}$ and $V_{THC}$ relative to low-conductance and high-conductance peaks, respectively.
c, $\varphi=0.87\,V_T$ is plotted as a function of force for $C_8$, $C_{12}$ and $C_{18}$ SAMs. Parameter *a* is extracted from these datas using Eq. 3.
d, Schematic picture proposed in ref.48, to denote the tilt-dependent molecular gate effect arising from changes in the effective interfacial dipole (vector sum of the red arrows), which has contributions from a permanent surface dipole and a molecular dipole oriented along the S-C bond. When the molecules are tilted by θ, the non-permanent molecular S-C dipole is reduced and the work function of the decorated surface increases. As sketched, this results in an upward shift of the molecular orbitals with respect to the Fermi energy, and hence in an enhanced tunnelling through the tail of HOMO resonance.
e, KFM CPD image for $C_8$ decorated Au nanocrystals.
f, CPD histograms (up) for $C_8$, $C_{12}$ and $C_{18}$ decorated Au nanocrystals. The increased tilt angle for short molecules tends to reduce the molecular dipole. When CPD is plotted as a function of the number of C, a linear dependence is obtained (slope 14 mV/C). We obtained CPD=220 mV for reference sample with uncoated Au nanocrystals (see SI).

related to the fact that $C_{18}$ SAMs are more deformed under strain than $C_8$ SAMs (Fig. 3a).

Considering the SAM deformations $\Delta d$ determined above (Eq. (1) and Fig. 3a), the force-dependent $\beta_{eff}$ is written as:

$$\beta_{eff} = \beta_0 \sqrt{\frac{\varphi}{\varphi_0}} \left[ 1 - \frac{\Delta d}{d_0} \right] \approx \beta_0 \sqrt{\frac{\varphi}{\varphi_0}} \left[ 1 - \frac{F}{SE_{SAM}} \right] \quad (2)$$

where $\beta_0$ and $\varphi_0$ are the tunnel decay ratio and the tunnel barrier height, respectively, when no force is applied on the SAM; and $\varphi(F)$ is the average force-dependent tunnel barrier (Fig. 4d, inset). In the first step, (considering only the monolayer deformation, i.e. $φ=φ_0$), a good fit is obtained (Fig. 4d) for F<10nN with $β_0$=1.05 per carbon atom, and using $E_{SAM}$=0.7GPa (in the range of measured values). $β_0$ is in agreement with values previously reported for alkyl chains (0.8-1.2 per carbon atom)[20-22,35,36] either for SAMs or single-molecule junctions and either for monothiol or dithiol junctions, as discussed more extensively in review papers.[39,40] Above 10 nN, we observe (Fig. 4d) a deviation of $β_{eff}$ from linear dependence, which can be ascribed to an additional effect such as a possible dependence of the tunnel energy barrier on force. To check this issue, the transition voltage spectroscopy (TVS) method is used to determine the energy position of the molecular orbital in the junction.[42-46]

| Molecule | θ (deg) | F (nN) =(1-cosθ)$E_{SAM}$.S | $E_{HOMO}$ | Δφ (eV)* | Theoretical a (eV/nN) from DFT | Experimental a (eV/nN) from TVS |
|---|---|---|---|---|---|---|
| $C_8$ | 0 | 0 | -3.05 | 0 | 0.077 (0.019-0.108) | 0.03 |
|  | 26 | 4 | -2.81 | -0.24 |  |  |
|  | 35 | 7 | -2.44 | -0.61 |  |  |
| $C_{12}$ | 0 | 0 | -2.9 | 0 | 0.062 (0.016-0.088) | 0.014 |
|  | 30 | 5 | -2.54 | -0.36 |  |  |
|  | 50 | 14 | -2.08 | -0.82 |  |  |

**Table 1: Comparison of the force coefficient a obtained from DFT and experiments.**
Variation of $E_{HOMO}$ (the energy of the HOMO referred to the Fermi energy of the substrate) versus the tilt angle taken from references 48,49 for $C_8$ and $C_{12}$ dithiol molecules. Here we use the relation $F=(1-cosθ).E_{SAM}.S$ with $E_{SAM}$=0.7 GPa, corresponding to the value obtained from β-F curves (Fig. 3d). Δφ gives the movement of $E_{HOMO}$ with respect to the reference at θ=0. From the calculated slope of Δφ(F) we obtain a theoretical estimation of the parameter a in Eq.(3), [we also show the corresponding values if $E_{SAM}$ is allowed to change within the experimental estimates 0.5-2.8 GPa] in the range of the experimental results obtained by TVS (Fig. 5c).

*TVS technique to discuss force-induced HOMO level shift*

In this method, the energy barrier height, namely the energy offset φ (Fig. 4d, inset) between the Fermi energy of the metal electrode and one of the molecular orbitals, is estimated from the current-voltage (*I-V*) measurement (see Methods). In the interpretation of electron transport through a tunnelling barrier, the voltage at which a minimum is observed in this plot represents the transition voltage $V_T$ between the direct and Fowler-Nordheim tunnelling regime. In the case of molecular junctions, $V_T$ can estimate the energy position of the molecular orbital (relative to the Fermi energy of the electrodes) involved in the transport mechanism (here, we suppose that for Au/alkylthiol junctions the HOMO level[35,36] dominates transport, see the discussion below), via a simple relationship φ = $αV_T$, with α ~ 0.87 for symmetric barriers.[47]

We performed direct spectroscopic *I-V* measurements on an alkylthiolated nanocrystal with $C_8$ molecules (C-AFM tip at a stationary point contact on the nanodot junctions, see Methods) representative of each conductance peak (i.e., measured on nanodot molecular junctions belonging to the maximum of each peak). Replotting *I-V* curves (Fig. 5a) as TVS plots (Fig. 5b), we get the $V_{TLC}$ and $V_{THC}$ for the low-conductance and high-conductance peaks, respectively. The Pt top and Au bottom electrodes work functions are in the same range and do not induce significant asymmetry (see SI, Fig. S9). The results of φ ~ 0.87 $V_T$ are shown in Fig.5c for $C_8$, $C_{12}$ and $C_{18}$ molecules at forces up to 30 nN. For all nanodot junctions, the φ values at low force are in the range of the previously reported values for alkylthiol junctions on Au (1-1.9 V).[35-42] We observe a linear dependence of φ with the applied force

$$φ = φ_0 - aF \qquad (3)$$

with $φ_0 = 1.25 ± 0.15$ eV and a~0.013-0.03 eV/nN (Fig.5c). When we use Eq. (3) with averaged "a" in Eq. (2), we obtain a good fit of the whole $β_{eff}$ vs. force curve (Fig. 4d, other parameters in Eq. (2) unchanged).

*Discussion on electronic properties and interfacial dipoles*

Such a large force modulation of $β_{eff}$ and the position of the HOMO level in molecular junctions were not previously observed from C-AFM measurements for SAMs with large lateral extension on Au substrate electrode. These previous results are puzzling and contradictory, showing that $β_{eff}$ is almost constant[48,49] or slightly increasing[50] or decreasing[51] with the C-AFM loading force. This difference can come from several reasons: i) contrary to our case, the contact area is increased with force and consequently the force per surface unit is not constant; ii) these previous experiments used SAMs on polycrystalline evaporated Au, thus the Au/alkylthiol interface may have hindered the behaviour reported in our work.

Studies based on DFT calculations also show that $E_{HOMO}$ (the energy of the HOMO level referred to the Fermi level of the substrate) increases and $\varphi$ decreases with increased tilt angle θ (Table 1) due to the interfacial dipole (Fig.5d) that is expected to be located between the sulfur and first carbon atom.[52,53] Moreover, $\varphi(F)$, derived from $\varphi(\theta)$ (see Table 1), shows a linear decrease with the applied force and $a$, extracted from Eq. (3), is in agreement with the experimental results. As a simple picture, the tilt angle increases with load while the interfacial dipole, is reduced. The large stress arising at the NP interface below the SAM (320 MPa @ 10 nN, Fig. 2f) could also play a role in φ(F) through a modification of NP's work function.[54,55] However, considering a typical bandgap pressure coefficient for NPs (<10 mV/100 MPa),[54,55] estimated $\Delta\varphi$, due to the sole Au NP work function, would have been lower than 15% of the measured effect.

The $\varphi$-$F$ results obtained on alkylthiolated NPs from $\beta_{eff}$-$F$, using the TVS technique, and DFT calculations converge to a value of $a \sim 0.025$ eV/nN (Eq. 3). This result is one order of magnitude larger than the value reported for molecular junctions on planar substrate.[41] In addition, the proposed mechanism, tilt-dependent interfacial S-C dipole projection perpendicular to the NP interface, also differs from the previous suggestion[41] of thickness-dependent field effect, such as image-charge.

*Interfacial dipoles investigated by Kelvin probe force microscopy*

We further examine the role of interfacial dipoles in thiolated NPs by considering the reduction of $\varphi$ with the alkyl chain length at low force (Fig. 5c). Smaller alkylthiol SAMs have a larger tilt angle because the van der Waals forces between the molecules are reduced.[56] This observation is compatible with the interfacial dipole hypothesis. Kelvin probe force microscopy (see Methods and SI, Fig. S10) image obtained on $C_8$ is shown in Fig. 5e. From similar measurements on $C_{12}$ and $C_{18}$, we build contact potential difference (CPD) histograms (between the SAM and the tip). They reveal that the CPD measured on the alkylthiolated nanocrystals increases linearly (14 mV/C) with chain length (Fig. 5f). This feature corresponds to a decrease of the work function of the alkylthiolated Au ($W_{Au}$) when increasing chain length. This result is compatible with the dipolar representation in Fig. 5d. From the Helmoltz equation (see SI), we deduce that the perpendicular projection of the Au/SAM dipole ($\mu_z$) decrease from ~0.6 D (for $C_{18}$) to ~0.35 D (for $C_8$). The smaller the chain length, the higher the tilt angle and smaller S-C $\mu_z$. The CPD results on thiolated NPs are in the same range as those obtained on thiolated gold substrates,[57] and are also in par with Ultraviolet Photoelectron spectroscopy (UPS) measurements.[58] The force-dependent experiments presented here, together with DFT simulations and KFM experiment suggest that charge transport occurs through HOMO level (Fig.5d). In fact, an increase of $W_{Au}$, when decreasing chain length, would correspond to an increase of the energy barrier height with the LUMO and a decrease of the energy barrier with the HOMO (Fig.5d). According to our TVS results (Fig.5c), we can conclude to a HOMO mediated transport in our case.

## Conclusions

Here, we described the elastic and electronic properties of alkylthiolated gold nanocrystals. We find that the estimated Young's modulus of pure facetted gold NPs is four times smaller than the usually considered bulk modulus, which could be explained with the recent suggestion[16] that anisotropy in elastic properties should be considered for single-crystal NPs, the presence of twins in our NPs and the large surface to volume ratio. We also estimated the Young's modulus of the alkylthiol monolayer to be ~ 1.4 GPa, by combining AFM measurements of the monolayer deformation and FEA simulations. This value is consistent with the results of structural relaxations based on DFT calculations that estimate the Young's modulus of the layer as a function of the rotational and twist angles. The nanoscale molecular junctions formed by these alkylthiolated nanocrystals contacted by a CAFM tip show strong decreases of the tunnel decay constant $\beta(F)$ and of the effective potential barrier height $\varphi(F)$ as function of applied force even in the few nN regime. Combined with FEA and ab-initio calculations, these results are satisfactorily explained by the strain-induced molecular deformation and the strong impact of the interfacial dipole on the molecular orbital position. This study at the single nanocrystal level provides a reference on a model system for the elastic and electronic properties of NPs, important for various NP-based applications such as strain gauges[59] and self-powered triboelectric sensors.[60] As SAMs Young modulus is not expected to change significantly with complexity, these results should be partly transposed to different organic coatings with consideration of lower $\beta_0$ for π-conjugated oligomers. In addition, linkers composed of a thiol bond and a short alkyl chain are often part of more complex molecules, including biomolecules, which suggests a similar contribution from interfacial dipoles. These findings show that even small van der Waals interactions in the nN range,[61] for example between NPs or between NPs and carbon nanotubes[62] or graphene, could be sufficient to alter the electronic properties of a wide variety of NP-based molecular devices. Similar measurements would be of great interest to other "functional" molecular junctions, such as "mechanical" switches (diaryethene, azobenzene) for which the applied force may also impact the isomerization, and thus the electrical conductance switching of these molecular devices.

## Acknowledgements

The authors would like to thank R. Arinero from IES Montpellier, T. Melin from IEMN for fruitful comments, and S. Lamant, J. Oden for assistance in FEA. P.L. is Senior Research Associate from the Fund for Scientific Research of Belgium (F.R.S. – FNRS). K.S. has been supported by Nord-Pas-de Calais Council fund and ANR project SAGE III-V (n°ANR11BS1001203) and S.D. by EU project I-ONE (FP7 n°280772). Experiments were partly funded by the SINGLEMOL project supported by Nord-Pas-de Calais council fund.

## Notes and references

[a] Institute of Electronics, Microelectronics and Nanotechnology, CNRS, Avenue Poincaré, 59652, Villeneuve d'Ascq France

[b] Centro de Fisica de Materiales, Centro Mixto CSIC-UPV, Paseo Manuel de Lardizabal 5, Donostia-San Sebastian, Spain

[c] Donostia International Physics Center (DIPC), Paseo Manuel de Lardizabal 4, Donostia-San Sebastián, Spain

[d] IKERBASQUE, Basque Foundation for Science, E-48011, Bilbao, Spain

[e] Depto. de Física de Materiales UPV/EHU, Facultad de Química, Apdo. 1072, Donostia-San Sebastián, Spain

[f] Laboratory for Chemistry for Novel Materials, Center of Innovation and Research in Materials and Polymers (CIRMAP), University of Mons, UMONS, Place du Parc 20, 7000 Mons, Belgium

† **Methods:**

Self-Assembled Monolayers(SAMs).

For the SAM deposition, we exposed the freshly evaporated gold nanodots to 1mM solution of alkylthiols (from Aldrich) in ethanol (VLSI grade from Carlo Erba) during 15 h. Then, we rinsed the treated substrates with ethanol followed by a cleaning in an ultrasonic bath of chloroform (99% from Carlo Erba) during 1 min.

AFM (peak-force, C-AFM and KFM).

Peak-force AFM measurements were performed with the recently developed peak force tapping mode (PeakForce-Quantitative Nano-Mechanics). Silicon cantilevers (Bruker AXS ©) with a spring constant of 150-250 $N.m^{-1}$ were used for experiments on uncoated Au nanocrystals and of 0.1-0.3 $N.m^{-1}$ for coated Au nanocrystals. Cantilever spring constant and sensitivity were calibrated before and after each experiment. In the present experiment, we didn't use the DMT modulus package to directly obtain an image of the young modulus this model is not appropriate to our nanocrystals that have a dimension much smaller than tip curvature radius. As a consequence, we have selected the direct measurement of deformation that can be converted into a Young modulus by FEA. Data processing was performed using the commercial Nanoscope Analysis software (Bruker AXS ©) and Wsxm (Nanotec.es).[63]

We performed current voltage measurements by conducting atomic force microscopy (C-AFM) in $N_2$ atmosphere (Dimension 3100, Veeco), using a PtIr coated tip (same tip for all C-AFM measurements). Tip curvature radius is about 40 nm (estimated by SEM), and the force constant is in the range 0.17-0.2 N/m. The conductance of the Au nanodot without molecule is much larger than that for Au nanodots with molecules and, in that case, dots are often burnt after/during such measurements probably due to the large current density. In the scanning mode, the bias is fixed and the tip sweep frequency is set at 0.5 Hz. Since our experimental setup is limited to 512 pixels/image, it leads to a typical number of counts of 2700 for a 6x6 μm C-AFM image. In the spectroscopy mode, representative molecular junctions belonging to each conductance peak are first identified from the C-AFM image. Because of imprecise positioning of the tip, 100 spectroscopic *I-V* curves are taken around this dot using a square grid (10 x 10 points with a lateral step of 2 nm). A significant current can only be measured when the tip is on top of the dot and thus a single *I-V* (with the maximum current) from these 100 *I-V* curves is selected per dot. TVS is obtained by plotting the *I-V* data in the form of a Fowler-Nordheim plot *(ln(I/V²))*.

The KFM measurements were conducted using a Dimension 3100 atomic force microscopy (AFM) system in a controlled Nitrogen environment glove box. We used Pt/Ir (0.95/0.05) metal-plated cantilevers with spring constant of ~3 N/m and a resonance frequency of ~70 kHz. First, the height profile was recorded in tapping mode. Then the potential or phase profiles were measured in noncontact lift mode at a height of 25 nm above the surface.

Nanodot histograms

We use our developed OriginC program for threshold analysis (given in Supporting Information). One count corresponds to the maximum current for one nanodot.

FEA

COMSOL v4.3 with Structure and deformation package was used to evaluate the Young modulus of both NP and SAMs. The design was performed in 2D-axisymmetry to simplify calculations. Tip and silicon substrate Young moduli of 170 GPa and 131 GPa, respectively, were considered.

DFT calculations

The monolayer was modeled using periodic boundary conditions with one molecule in each unit cell. The orientation of the molecule in the monolayer is defined by the three angles shown in Fig. 3b. A x 30° lattice was assumed with the experimental intermolecular distance d = 5.05 Å. Several structural relaxations were performed as explained in the Supplementary Information. First the tilt angle θ of the molecules was fixed to the commonly reported value of 30° and the relevant values of (Φ, ψ) were determined by exploring the energy landscape. Afterwards the energy versus tilt curve was obtained for several configurations with fixed values (Φ, ψ). We chose this approach to determine, on the one hand, the dependence of the results on the details of the structure and, on the other hand, to obtain smooth curves that would allow numerical differentiation to obtain an estimation of the F(θ) curves. The electronic structure calculations and the relaxations were performed with the DFT code SIESTA[64] using a real-space grid of 400 Ry and a double-ζ plus polarization (DZP) basis for the C, and H atoms with an energy shift of 0.02 Ry. We used a Brillouin zone sampling of 6 × 6 × 1 k-points. The height of the supercell in the z-direction was fixed to 25 Å for C8 and 30 Å for C12, so there is enough vacuum to avoid interactions among periodic replicas of the monolayers. Total energy was converged with respect to these parameters.

# On the Mechanical and Electronic Properties of Thiolated Gold Nanocrystals

K. Smaali[1,λ], S. Desbief[1,λ], G. Foti[2,3], T. Frederiksen[3,4], D. Sanchez-Portal[2,3], A. Arnau[2,3,5], J.P. Nys[1], P. Leclère[6], D. Vuillaume[1] and N. Clément[1]*

## Supplementary Information

[1]Institute of Electronics, Microelectronics and Nanotechnology, CNRS, Avenue Poincaré, 59652, Villeneuve d'Ascq France

[2]Centro de Fisica de Materiales, Centro Mixto CSIC-UPV, Paseo Manuel de Lardizabal 5, Donostia-San Sebastian, Spain

[3]Donostia International Physics Center (DIPC), Paseo Manuel de Lardizabal 4, Donostia-San Sebastián, Spain

[4]IKERBASQUE, Basque Foundation for Science, E-48011, Bilbao, Spain

[5]Depto. de Física de Materiales UPV/EHU, Facultad de Química, Apdo. 1072, Donostia-San Sebastián, Spain

[6]Laboratory for Chemistry for Novel Materials, Center of Innovation and Research in Materials and Polymers (CIRMAP), University of Mons, UMONS, Place du Parc 20, 7000 Mons, Belgium

λ: These authors contributed equally to the study

# 1 Deformation of the substrate and facets

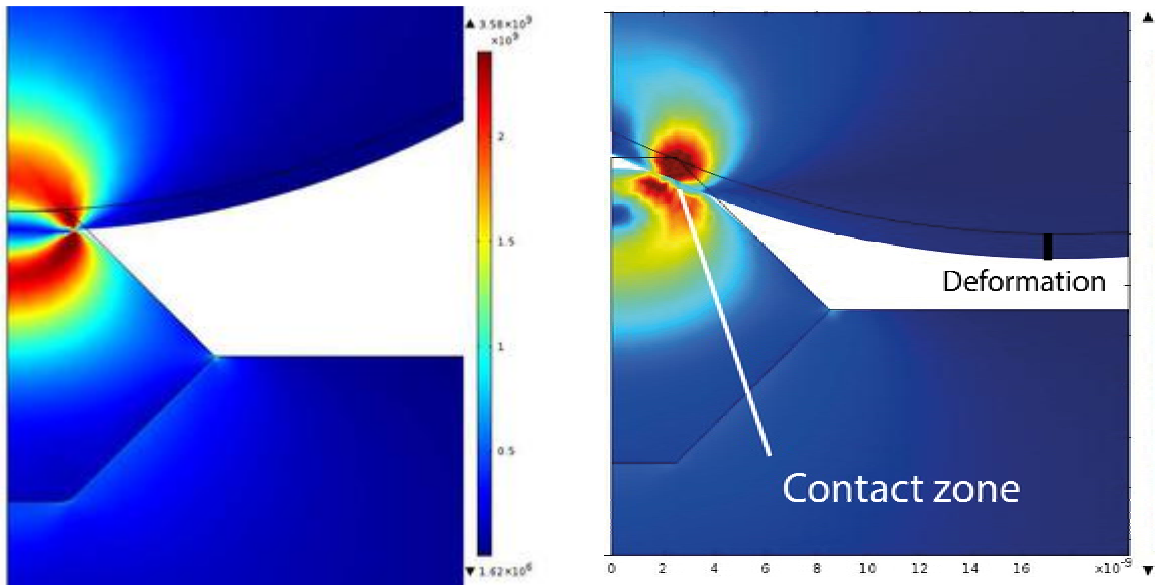

**Figure S1** Finite Element Analysis (FEA) illustrating the simulation of a deformation measurement with peakforce AFM when the tip is located above the nanodot (left) and above facets (right). Black lines represent the initial geometry of the system at zero force. We observe a negligible deformation of the substrate in both cases. On the right image, although the contact zone is located only on the edge, the measured deformation (larger than expected) is done at tip center.

## 2 Tip indentation: "nano" SAM vs large area SAM

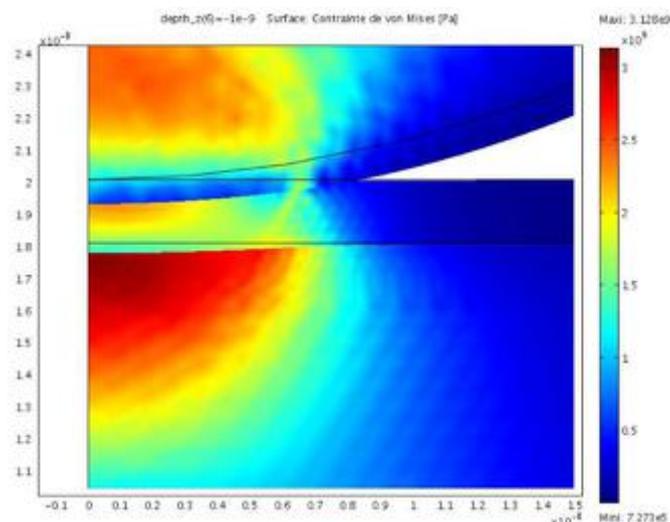

**Figure S2a** Like previous figure but showing an example of the simulation of the tip indentation in a conventional "large area" SAM. As the tip indents the SAM, the contact area increases.

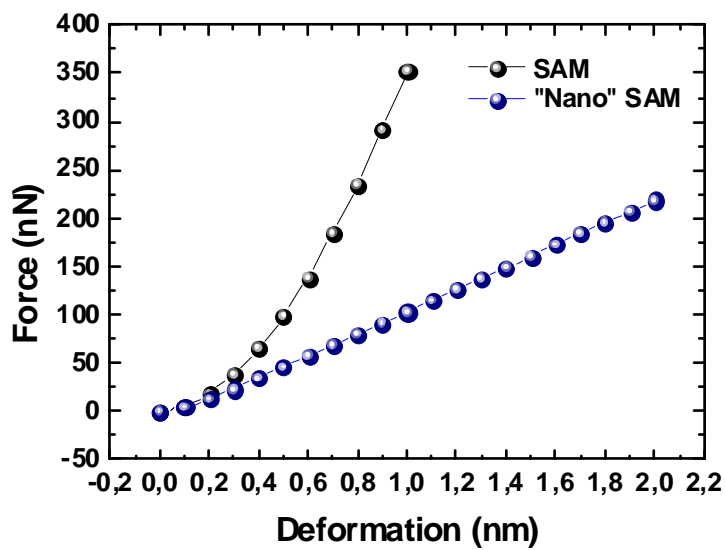

**Figure S2b** Force-deformation curve obtained by FEA (assuming a Young's modulus of the SAM of 5 GPa) for a "nano" SAM on a gold nanocrystal and a SAM. Whereas in the first case a linear effect is observed, a parabolic behavior is observed in the second case.

## 3 Adhesion mapping

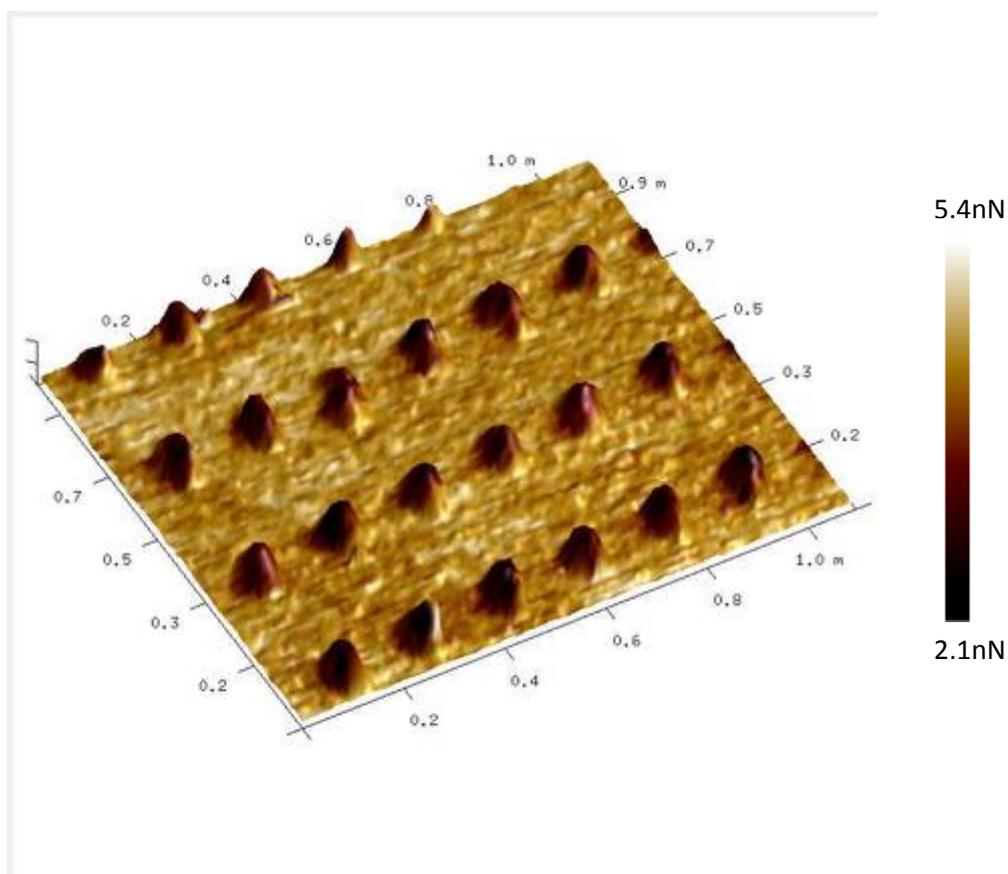

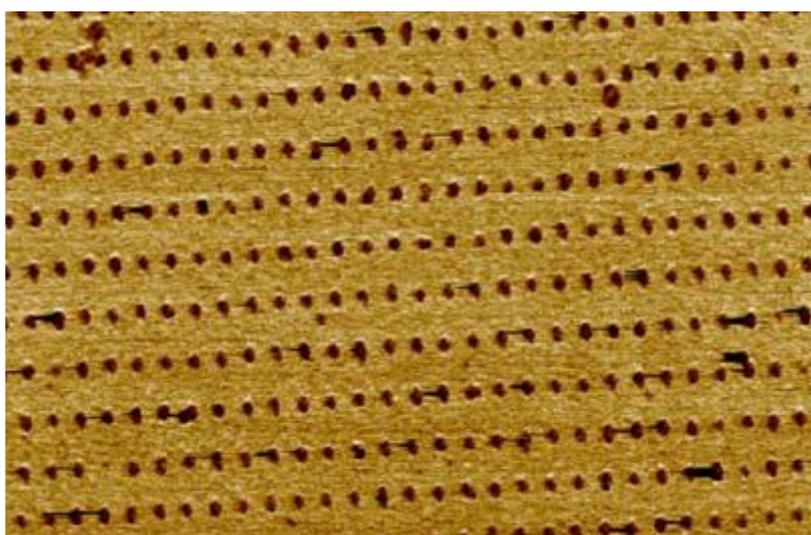

**Figure S3** Adhesion-topography coupled images (two different dimensions) of Au nanoparticle (NP) coated with $C_8$. The adhesion is strongly suppressed on top of the Au NP.

## 4 SAMs thicknesses

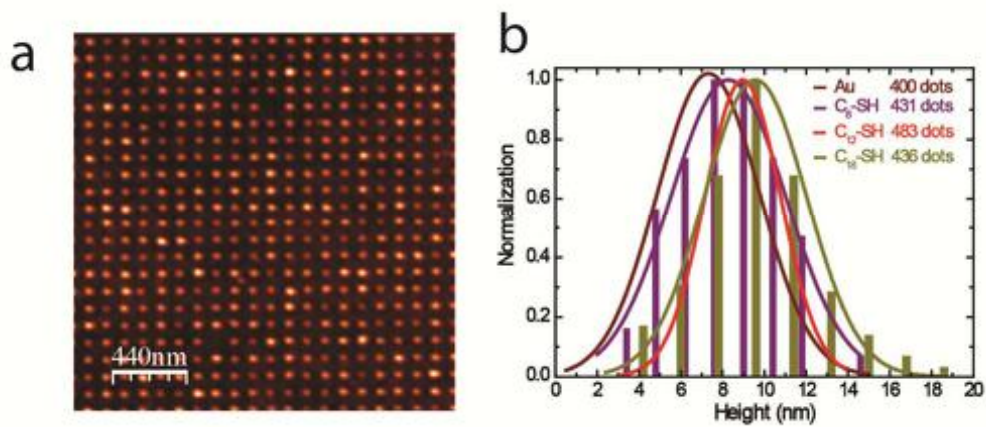

**Figure S4** (a) AFM image in tapping mode of an array of gold nanodots covered with C12 molecules (1 μm x 1 μm image with 1024 x 1024 pixels). (b) Normalized height histograms on nanodots and nanodots covered with $C_8$, $C_{12}$ and $C_{18}$ molecules.

## 5 Ab-initio density functional method

In order to determine the minimum energy orientation of $C_8$ and $C_{12}$ we calculated the total energy as a function of the twist and rotational angles respectively [Fig. S5 (a) and (b)] using the relaxed geometry of the isolated molecules assuming a tilt angle $\theta=30°$. Without a proper description of van der Waals (vdW) interactions it is difficult to obtain an accurate estimation of the tilt angle. Therefore, for this initial set of calculations we took this value from the literature, where it is frequently reported [1-3].

We found that the energy landscape shows a periodicity of 60° for Φ (due to the $C_6$ symmetry axis of the hexagonal lattice) and 180° ($C_2$ axis of the alkane chain) for Ψ [Fig. S5 (c)].

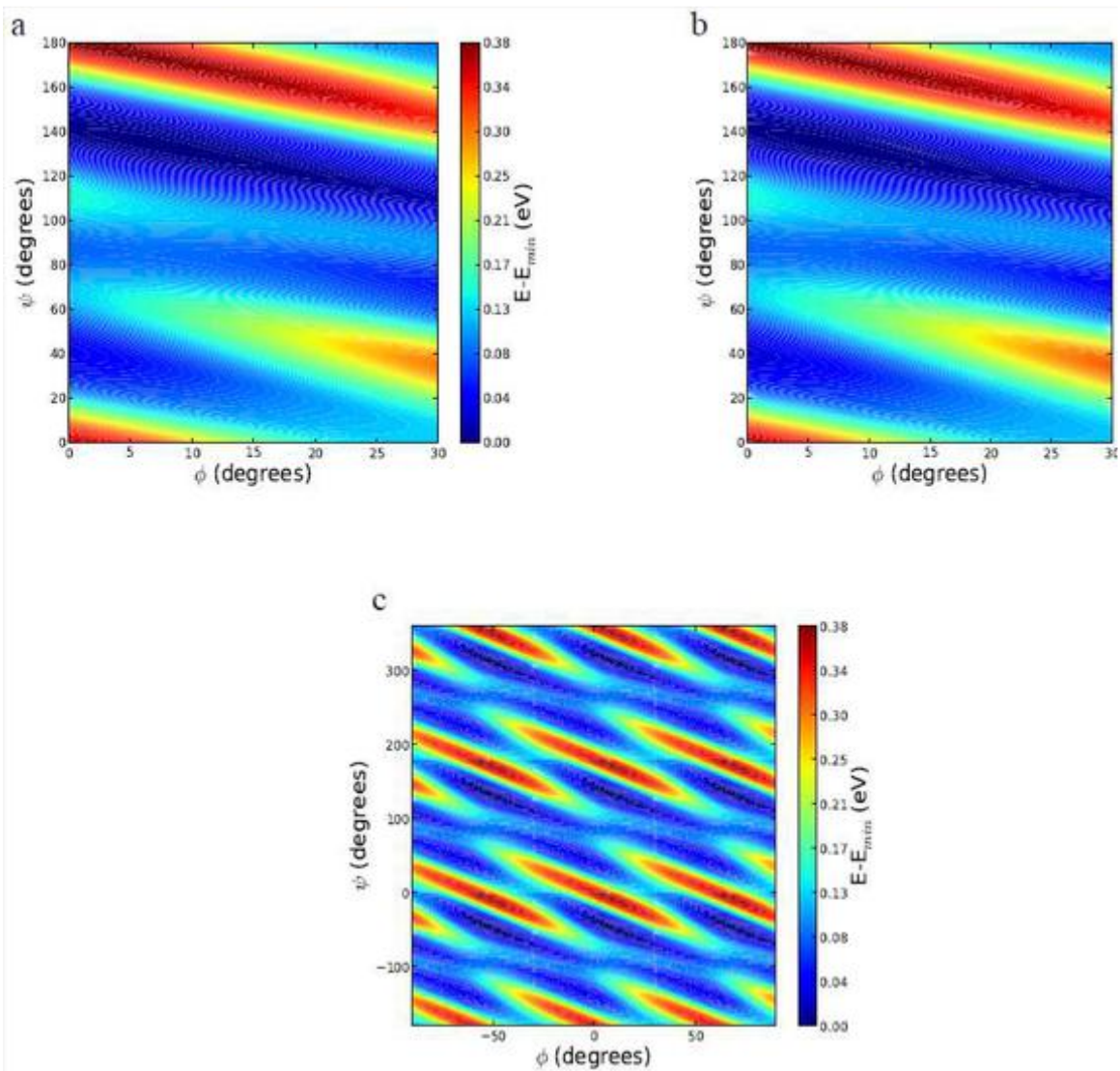

**Figure S5** Total energy as a function of twist and rotational angles for a) $C_8$ and b) $C_{12}$ with PBE functional. The tilt angle θ has been fixed to 30°. In the case of $C_8$, the minimum of energy is at Φ=9°, Ψ=132° while in the case of $C_{12}$ the minimum of energy is at Φ=8°, Ψ=132°. In panel c) total energy is represented as a periodic repetition of the unit cell defined in the interval: -30≤Φ≤30 and 0≤Ψ≤180.

When using Pedew-Burke-Ernzerhof (PBE)[4] functional we found that, in the case of $C_8$ the minimum of energy is at Φ = 9°, Ψ = 132° while in the case of $C_{12}$ the minimum of energy is at Φ = 8°, Ψ= 132°. Using a functional that includes vdW corrections[5] the two minima shift in both cases to Φ = 6° and Ψ= 138°. So we found that, if no effects of substrate and anchoring groups are considered and assuming the same coverage (same d=5.05 Å, distance between molecules), $C_8$ and $C_{12}$ relax to the same minimum energy configuration as already found in a previous work.[6] Note that Φ =30°, Ψ=90°, the configuration that gives a Young modulus $E_{sam}$ within the experimental range of values, is not far from a minimum of energy. After determining the relevant range for Φ and Ψ, we relaxed four different configurations corresponding to different combinations of Φ and Ψ. We did that using the z-matrix coordinates until residual forces were lower than 0.03 eV/Å and 0.003565 Ry/rad. Then, for each conformation, we tilted the molecule as a rigid rod with respect to the carbon atom of the lower $CH_3$ group in order to calculate the tilt-vs-force curve. The approximation of tilting the molecule rigidly is justified by two facts: i) the molecule is embedded in a monolayer with small room for bending, and ii) much larger stiffness of the molecules respect to stretching than respect to bending or rotation. We chose this approach of fixing (Φ,Ψ) in order to obtain smooth energy versus tilt angle curves that we can numerically differentiate to get the stress versus tilt. The result shown in Fig. 5 c and d in the paper indicate a similar elastic response of the monolayers formed by C8 and $C_{12}$ molecules.

The independence of the Young modulus on the length of the chain is, for an isolated chain constrained to keep a straight conformation, an expected result (at least for a sufficiently long chain). However, it is more interesting the fact that this length independence still holds for the close-packed layer of chains forming an angle with respect to the direction of the load, and this angle being a function of the applied stress.

Further insight could be provided by looking at the energy surface as a function of Φ and Ψ. For short and long chains we get basically the same potential energy landscape [Fig. S5 a) and b)]. Its derivative with respect to the tilt (which is the force) is the same. This means that, for fixed Φ and Ψ, we expect the same potential energy curve as a function of θ for both $C_8$ and $C_{12}$. Using the previous data we can get an estimation of Young's modulus $E_{SAM}$ as function of tilt. The Young's modulus is defined as (notice that this definition extends to finite deformations):

$E(\theta) = d\sigma/d\varepsilon$ (Eq.S1)
$= 1/A \cdot dF/d\varepsilon$
$= 1/A \cdot \cos(\theta_0)/\sin(\theta) \cdot dF/d\theta$

where σ is the stress expressed as a force F per unit area A while ε is the adimensional strain $\varepsilon = (z_0 - z)/z_0$ with respect to the equilibrium position $z_0 = L_0 \cos(\theta_0)$ and $L_0$ is the length of the molecule.

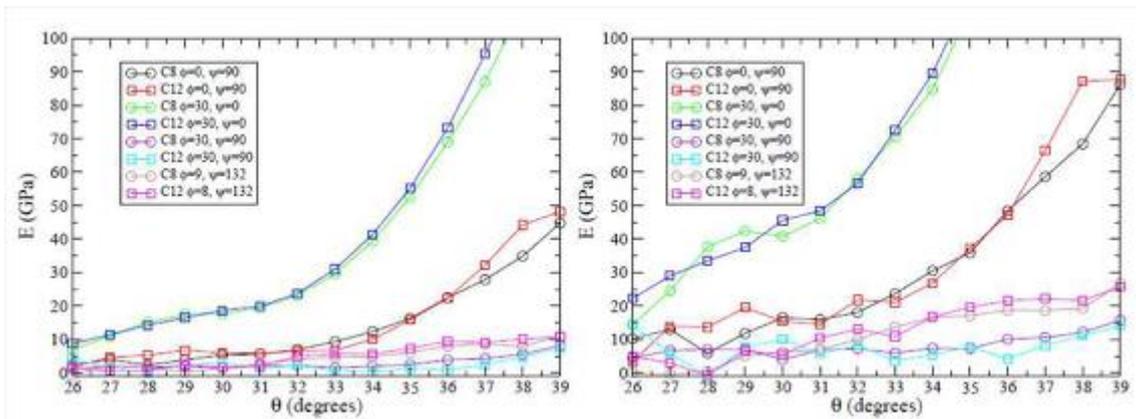

**Figure S6** Calculated Young's modulus E(θ) for $C_8$ and $C_{12}$ as a function of tilt angle using PBE (left) and vdW (right) functional comparison

As we can see from Fig. 3c in the paper Young's modulus depends not only on the tilt angle θ but also on the other two angles Ψ and Φ. When tilting the molecules, due to intermolecular interactions, $E_{SAM}$ increases with a slope which depends on the particular orientation of the molecule. There are no remarkable differences between PBE and vdW calculation. Young's modulus for the minimum energy configurations close to the equilibrium position (θ between 33° and 35°) is between 4.66 GPa and 19.4 GPa. We should stress that these values should be considered as an upper limit of the theoretical prediction. Firstly, since molecules are not fully relaxed for each angle θ, a steeper slope of the force with respect to the tilt is obtained. This means that our results have to be considered as an upper limit of the real values. Secondly, in our calculations we did not consider the effects of both surface and anchoring groups. In Fig. S6 we show the Young's modulus as a function of the angle when using PBE and vdW functional respectively for different (Φ,Ψ) configurations.

## 6 Normal distribution

If X is a random variable with a normal distribution, then Y=exp(X) has a log-normal distribution; likewise, if Y is log-normally distributed, then log(Y) is normally distributed.

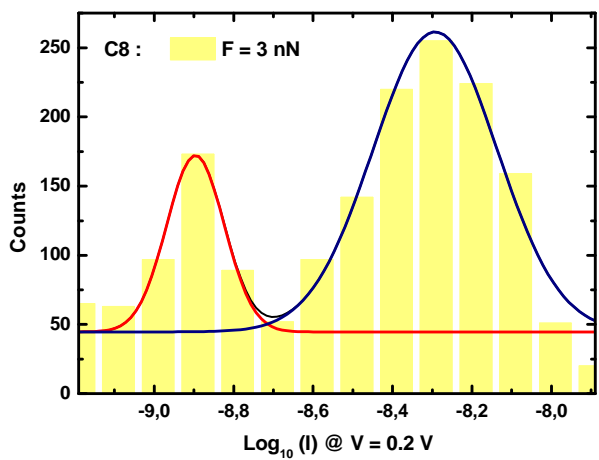

|  |  | F (nN) & V = + 0.2 V | | | | | |
|---|---|---|---|---|---|---|---|
|  |  | 3 | | 7.5 | | 30 | |
|  |  | Log$_{10}$ (I (A)) | FWHM | Log$_{10}$ (I (A)) | FWHM | Log$_{10}$ (I (A)) | FWHM |
| C8 | HC | -8.29 | 0.31 | -7.90 | 0.16 | -7.71 | 0.03 |
|  | LC | -8.89 | 0.14 | -8.13 | 0.12 | -7.79 | 0.17 |
| C12 | HC | -9.89 | 0.24 | -8.99 | 0.19 | -8.82 | 0.10 |
|  | LC | -10.37 | 0.34 | -9.41 | 0.36 | -8.55 | 0.17 |
| C18 | HC | - | - | -10.44 | 0.23 | -8.84 | 0.07 |
|  | LC | - | - | -10.99 | 0.26 | -8.94 | 0.02 |

**Figure S7**: Top: Log(I) histograms for $C_8$-coated NPs. The 2 peaks are fitted with log-normal functions. Bottom: Fits results for $C_8$, $C_{12}$, $C_{18}$ molecules and different forces are reported in the table.

## 7 Thermal effects

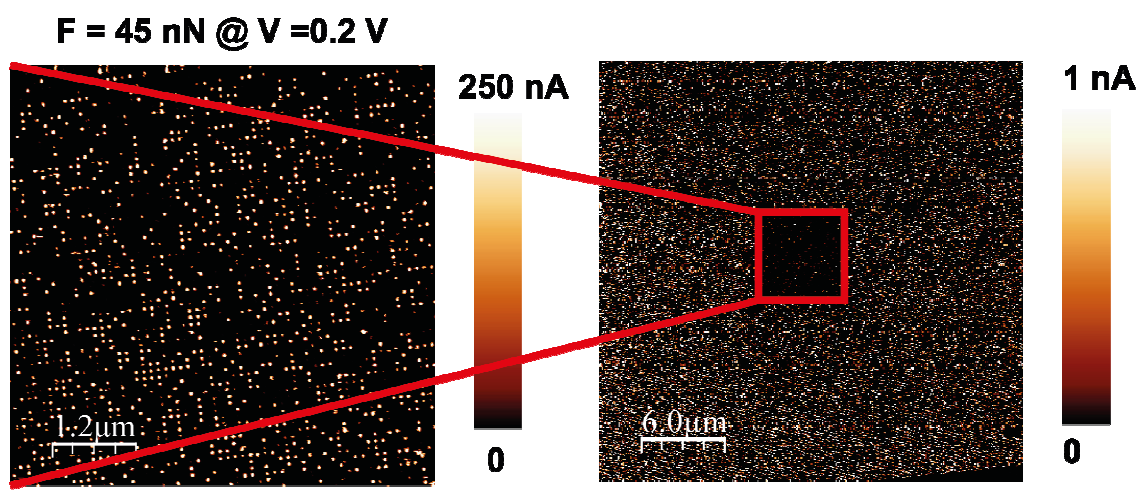

**Figure S8** Conducting Atomic Force Microscope (CAFM) image left when a force (~45 nN) and 0.2 V is applied on $C_8$-coated Au NP. As a result of coupled large force/large current density, sublimation of Au NPs is noticed in a second (enlarged) CAFM image.

## 8 Transition Voltage Spectroscopy

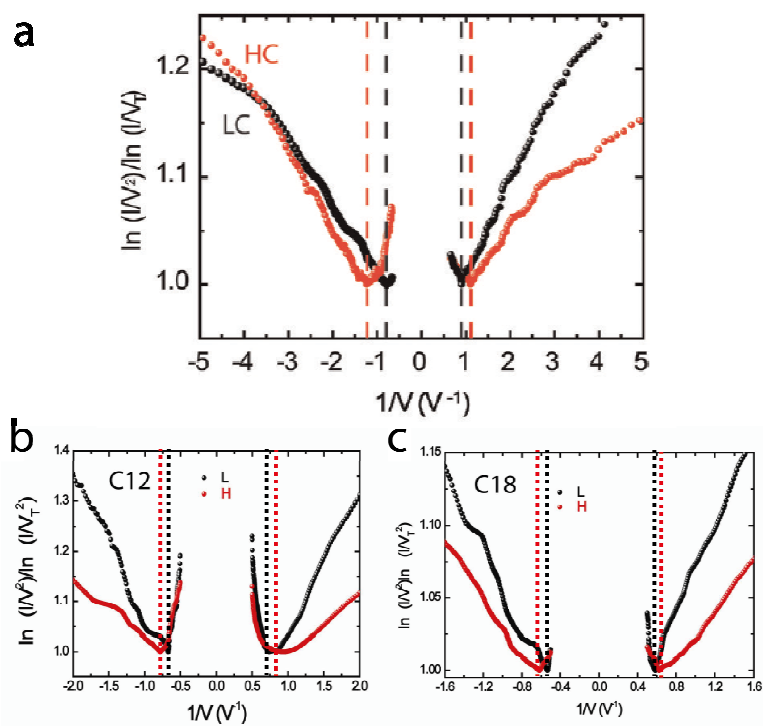

**Figure S9** Fowler-Nordheim plots for a) $C_8$-, b) $C_{12}$-, c) $C_{18}$- Au nanocrystals. $V_T$ (minimas in these plots) are symmetric in positive and negative voltages.

## 9 Kelvin Force Microscopy (KFM)

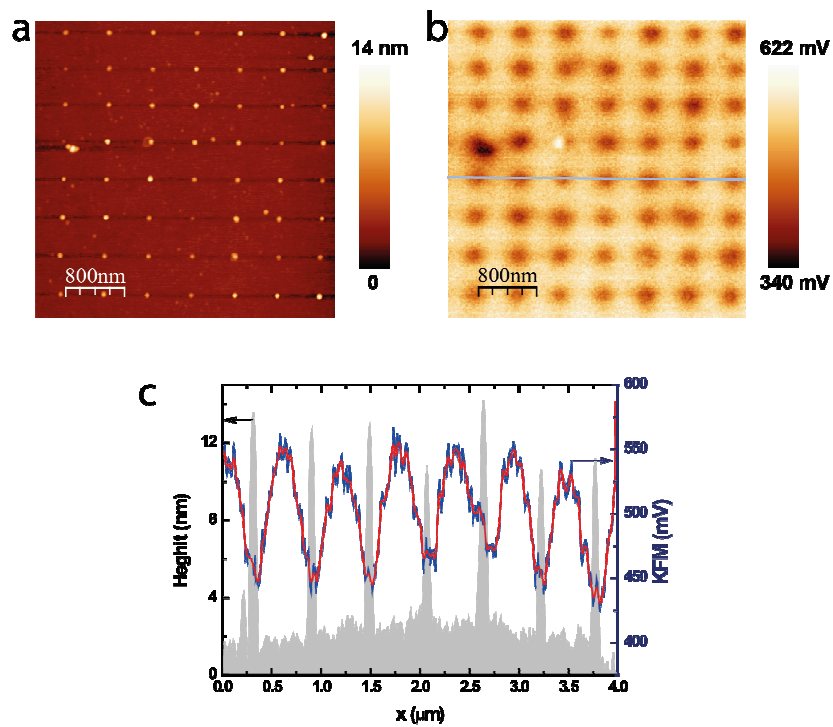

**Figure S10** Topographic (a), Kelvin Force Microscope (KFM) (b) and cross section views of both topographic and KFM images (c) for $C_8$-coated Au NPs.

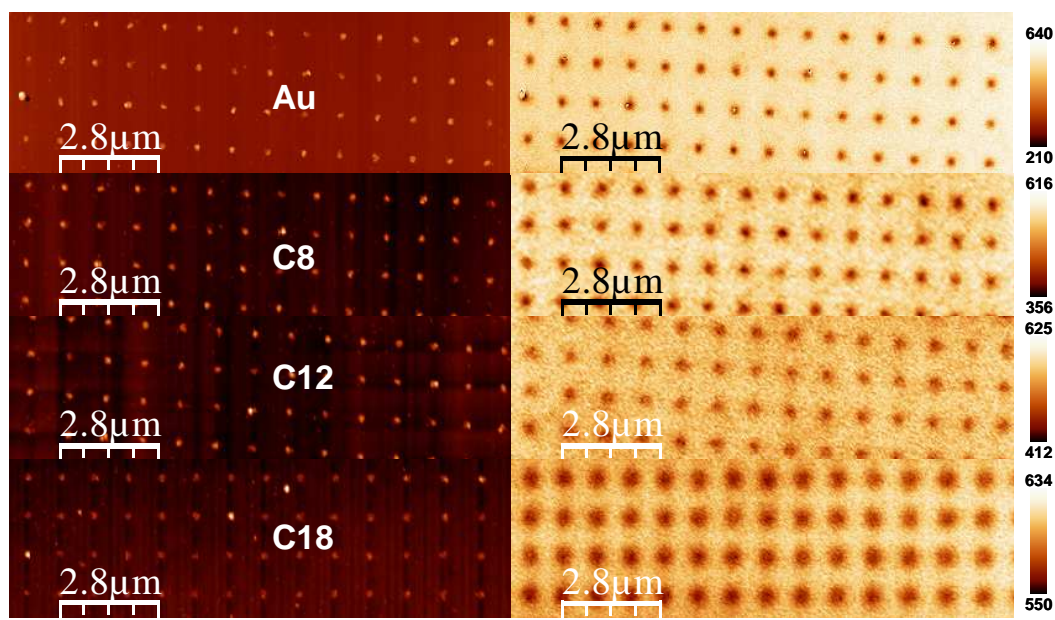

**Figure S11** Topographic (left) and KFM (right) images for naked, $C_8$-, $C_{12}$-, $C_{18}$-Au NPs.

# 10 Origin C program used for the treatment of C-AFM images on the array of nanodots

The 1st function applies a threshold to put 0 in the matrix below the threshold (removal the background noise). Then, the maximum per dot is obtained by checking the nearest neighbors (function maxi).

```
void Threshold(string strName, double thmin, double thmax, int ibegin, int iend, int jbegin, int jend)
{
        Matrix mm(strName);
        for (int i=ibegin; i<iend; i++)
                for (int j=jbegin; j<jend; j++)
                        if ((mm[i][j]<thmin)||(mm[i][j]>thmax)){mm[i][j]=0};
}

void maxi(string strName, int neighbors, int ibegin, int iend, int jbegin, int jend)
{
        Matrix mm(strName);
        for (int i=ibegin; i<iend; i++)
        {
                for (int j=jbegin; j<jend; j++)
                {
                        if(mm[i][j]!=0)
                        {
                                for (int k=¶ 1*neighbors; k<=neighborss; k++)
                                {
                                        for (int l=¶ 1*neighbors; l<=neighbors; l++)
                                        {
                                                if (((i+k)>=0)&&((j+l)>=0)&&((i+k)<iend)&&
                                                ((j+l)<iend))
                                                if(mm[i+k][j+l]>mm[i][j]) {mm[i][j]=0};
                                        }
                                }
                        }
                }
        }
        int a=0;
        Worksheet wks;
        wks.Create("histogram.otw");
        WorksheetPage wksp=wks.GetPage();
        wksp.Rename("histogram");
        string str;
        for (int m=ibegin; m<iend; m++)
        {
                for (int n=jbegin; n<jend; n++)
                {
                        if (mm[m][n]!=0)
                        {
                                str.Format("%f",mm[m][n]);
                                wks.SetCell(a, 0, str); // set the value to a cell of worksheet
                                a++;
                        }
                }
        }
}
```

Program call in Origin Labtalk window :
Threshold(Matrixname,threshold_min_nb,threshold_max_nb,0,8192,0,8192);
maxi(Matrixname,nb_neighbors,0,8192,0,8192);
Typically, we use 5 neighbors.

**11 Interface dipole**

The perpendicular projection of the interface dipole, $\mu_z$, is given by the Helmholtz equation:

$$\Delta CPD = \frac{N\mu_z}{\varepsilon_0 \varepsilon_{SAM}}$$

where $\Delta CPD = CPD_{SAM} - CPD_{Au}$, $N$ is the surface density of molecules in the SAM, $\varepsilon_0$ is the vacuum dielectric permittivity, $\varepsilon_{SAM}$ is the relative permittivity of the SAM. From KFM on naked Au NPs, we have $CPD_{Au} \sim 220$ mV. We chose an average value of $4\times10^{14}$ cm$^{-2}$ for $N$, assuming a reasonable molecule packing in the SAMs, and $\varepsilon_{SAM}$ =2.5. From the CPD values shown in Fig. 5f, we get $\mu_z$ = 0.58D, 0.37D and 0.34D for the $C_{18}$, $C_{12}$ and $C_8$ SAMs, respectively.

**12 Resicope image**

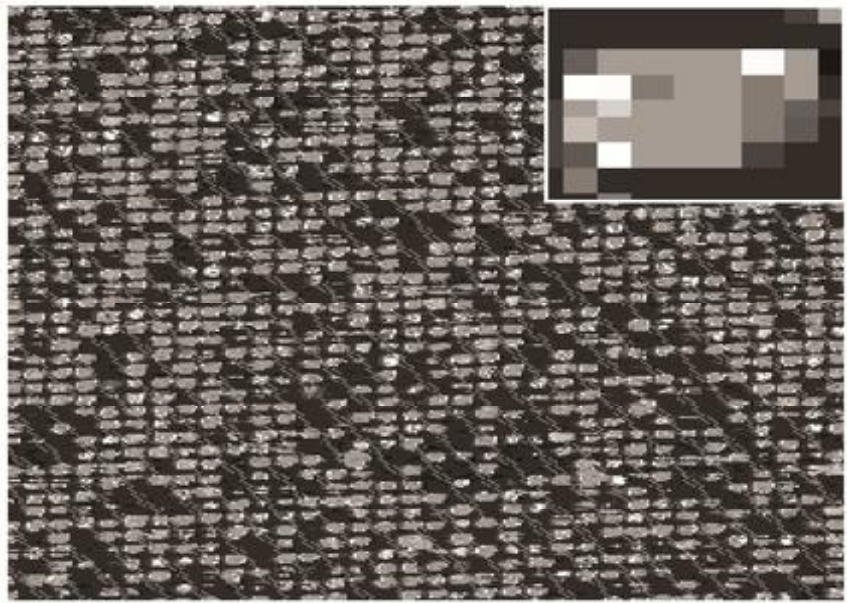

**Figure S12. 3.4 µm x 4.3 µm R-AFM image**

Resiscope-AFM image (log amplifier) of $C_{12}$ molecular junctions with crystal Au nanodot electrodes. Due to the high scan speed (10 µm/s) parasitic high current levels appear at dot borders (example: white pixels in the dot shown in inset) and the apparent tip curvature radius increased (distance between dots reduced). After thiolation and sample cleaning in the ultrasonic bath, 80-85% of the dots are still there.